# On Science, pseudoscience and String theory

Asis Kumar Chaudhuri
Variable Energy Cyclotron Centre
1-AF Bidhan Nagar, Kolkata-700 064

**Abstract:** The article discusses the demarcation problem; how to distinguish between science and pseudoscience. It then examines the string theory under various demarcation criteria to conclude that string theory cannot be considered as science.

## 1. Introduction

Which human activities can be considered as science, and which are not, is an issue of debate in the philosophy of science. It is known as the problem of demarcation; how to distinguish between science and its look alike or non-science or pseudoscience [1,2,3]. For example, astronomy has its origin in astrology, and throughout the world, both astronomy and astrology are practiced with most fervent. Yet, we unanimously agree that astronomy is a science but few will agree that astrology is also a science. On what basis we are drawing the distinction? Not only astrology, throughout the world, humans are engaged in several practices; phrenology (study of skull structure believing its relation with mental faculties), palmistry (study of lines on the palm hoping to predict the future), numerology (study of occult significance of numbers), iridology (study of the iris of the eye for indications of bodily health and disease), dowsing ( a type of divination employed in attempts to locate ground water, buried metals or ores, gemstones, oil, gravesites and many other objects and materials without the use of scientific apparatus), creationism (the religious belief in biblical interpretation of Universe and life), divination (the practice of using signs such as arrangements of cards or tea leaves, or special power to predict the future),  and many more. Are they science? Should the society encourage these types of practices? Will these practices, in the long run be beneficial or detrimental to the human progress?

The issue of demarcation between science and non-science or pseudoscience has an added significance today.  Some of the World's brilliant minds are pursuing the so-called string theory which has been hyped as the theory of everything. Yet, in recent years, a question was raised; is string theory science?  Indeed, there are ample reasons to raise the question. In the following, I will discuss the demarcation problem and several demarcation criteria in details and examine the string theory under the lens of the demarcation criteria. Unfortunately, it will be concluded that the string theory cannot be classified as science.

## 2. The demarcation problem

It is not easy to distinguish between science and pseudoscience and for long, scientists and philosophers have debated over the demarcation problem. One may ask; what constitute science? Indeed, if we know what science is, then possibly we can distinguish what is look-alike non-science or pseudoscience. The purpose of science is to develop general laws that explain how the world around us works and why things happen the way they do. For Aristotle, Science is the cultivation of apodictic knowledge; human knowledge characterized by evidence and certainty. The first principles of nature are directly intuited from sense and what we call science directly follow from these first principles. It is apodictic certainty that distinguishes science from other kinds of beliefs. Auguste Comte founded his 'doctrine of positivism' based on Aristotle's dictum. In the positivistic approach to science, experiences furnish us with particular facts and from the particular facts, using inductive logic, scientists find universal truth or truths. Induction is the process of inference when we draw universal statements or scientific laws on the basis of singular or particular facts or statements. Here is an example of use of inductive logic;

> Ram is a man.
> Ram is mortal.
> Jadu is a man.
> Jadu is mortal.
> Therefore, all men are mortal.

Induction relies on two fundamental principles; (i) the Law of Uniformity of Nature and (ii) the Law of Causation. Law of Uniformity of nature can also be expressed as "the future resembles the past" or "nature repeats itself." Thus, if in the past, under certain conditions, a particular phenomenon happened, then in future also, under the same conditions, the same phenomenon will happen. For example, if in the past, water quenches your thirst, in future also, water will quench your thirst. Law of uniformity of nature does not mean that there is a single law of nature governing all the aspects rather different aspects of nature are governed by different laws. The second principle, "the law of causation" expresses the causal relation between cause and effect, i.e. every event has a cause. One can be more precise about the causation principle; "Every phenomenon which has a *beginning* must have a cause."

However, is it justified to draw a universal statement from a particular statement or particular statements? While many believe that inductive process is the soul of the modern science, philosophers like Karl Popper and David Hume believed to the contrary [3,4]. In inductive logic, the unique character of science is 'verification': science deals with results, theories or experiments which can be verified. According to Karl Popper and David Hume, experiences, how numerous may be, do not allow you to draw a universal statement. They cite the famous example of 'black swan'. Till European explorers discovered Australia, there was unanimous agreement that all swans are white. But the age-old believe crumbled as soon as one black swan was discovered in Australia.

Karl Popper [3] in 'The Logic of Scientific Discovery' argued that verification cannot be a criterion of science. He argued that rather than verification, falsifiability should be the criterion of science. He wrote,

*"In so far as a scientific statement speaks about reality, it must be falsifiable: and in so far as it is not falsifiable, it does not speak about reality."*

Indeed, one can never conceive all possible situations for verification; on the other hand, a single instance of falsification will negate any theory based on numerous experiences or observations. According to Popper, theories are scientific if they can be falsified; if they are open to refutation. Non-scientific conjectures, theories, views cannot be refuted. The approach is more akin to deductive logic. In deductive logic, one starts with an assertion, with a given whole and infers from it the qualities of its parts. For example;

All men are mortal.
Ram is a Man.
Therefore, Ram is mortal.

One counter example will negate the assertion or the whole. Popper cites one classical example; the difference between astronomy and astrology. Astronomical theories, predictions or hypotheses may be proved wrong. But astrological theories, predictions cannot be proved wrong. Practitioners of astrology can always make suitable 'ad hoc' adjustments to suit the situation. They will be at liberty to say that their predictions failed for a specific person for so and so reasons, but it remained valid for some others.

Popper's demarcation criterion has been refuted by the noted philosopher Thomas Kuhn [5]. Kuhn argued that there are two types of science: normal science and revolutionary science. Normal science is the science conducted by the practitioners with an accepted paradigm of Worldview, i.e. they broadly agree about the way the world is. In revolutionary science, the paradigms are under attack and are subjected to change. For example, until Nikolas Copernicus, astronomers were engaged in normal science, with the accepted paradigm of geocentric or Earth centered model of Universe. Copernicus attacked the accepted paradigm and a period of revolutionary science followed. The accepted paradigm was changed and astronomers were back to normal science with a new paradigm of heliocentric of Sun centered Universe. Similarly, Einstein introduced a paradigm change in our understanding of space-time. Before Einstein's special Relativity, the paradigm was that space and time are separate entities. Einstein changed the paradigm to single entity space-time. Most of the time, practitioners of science are engaged in normal science, when within a given paradigm, scientists are confronted with anomalies or puzzles, which they try to solve. Occasionally, the normal science is interspersed by revolutionary science. In "The Structure of Scientific Revolutions" Kuhn wrote:

*"...no theory ever solves all the puzzles with which it is confronted at a given time; nor are the solutions already achieved often perfect. On the contrary, it is just the incompleteness and imperfection of the existing data-theory fit that, at any time, define many of the puzzles that characterize normal science. If any and every failure to fit were ground for theory rejection, all theories ought to be rejected at all times."*

For Kuhn, normal science is like puzzle solving, an activity aiming to solve scientific problems generated within a certain paradigm. When the problems are too large to be accommodated within the accepted paradigm, revolutionary science emerges. According to Kuhn, Popper's falsification criterion can be applied to revolutionary science only, but not to the normal science. What is the difference between science and pseudoscience? According to Kuhn, in science, at any time, in a given paradigm, there are several puzzles. But, in pseudoscience, never there is any puzzle.

We find that there is no unanimity among the philosophers over the demarcation problem. Indeed, few believe that it is not possible to demarcate science and non-science. Thus, Larry Laudan [1], in his influential essay, "Demise of the demarcation problem" critically examined several demarcation criteria and concluded that they do not serve the very purpose of distinguishing between science and non-science. He argued that the existing demarcation criteria do not provide a single necessary and sufficient condition for the demarcation problem. In "Demise of the demarcation problem" he wrote;

*"I will not pretend to be able to prove that there is no conceivable philosophical reconstruction of our intuitive distinction between the scientific*

*and the non-scientific. I do believe, though, that we are warranted in saying that none of the criteria which have been offered thus far promises to explicate the distinction."*

He termed the demarcation problem a pseudo-problem and wanted to remove terms like pseudoscience or unscientific from our vocabulary. According to him "... *they are just hollow phrases which do only emotive work for us."*

Even though Laudan termed the demarcation problem a pseudo-problem he argued to retain the distinction between reliable knowledge and unreliable knowledge. The rubric of reliable knowledge will include much which we commonly regard as scientific and exclude much that we commonly regard as non-science or pseudoscience. In a sense, Laudan did not kill the demarcation problem, rather rephrase it as the demarcation problem between reliable knowledge and unreliable knowledge.

The present author is of the opinion that demarcation problem is a complex problem and a single criterion cannot distinguish between science and pseudoscience. An activity, to be deemed as science needs to satisfy a set of criteria or characteristics. In his opinion, the following three criteria;

(i) **Positivistic verifications**: theories or statements can be verified;
(ii) **Karl Popper's falsification**: theories or statements can be falsified;
(iii) **Kuhn's puzzle-solving**: within a given paradigm, existence of anomaly or puzzle at any time;

will suffice to distinguish between science and pseudoscience. For example, all the three criteria will agree that astrology is a pseudoscience. It cannot be verified, nor it can be falsified, nor there is any puzzle in astrology.

### 3. Basics of String Theory

Let us discuss briefly about the string theory. An excellent introduction to string theory can be found in [6]. Originally, string theory was proposed as a theory for hadrons. In 1950-60's with the advent of particle accelerators, physicists faced the so-called problem of "Zoo of particles". Hundreds of particles were discovered which were considered to be fundamental in the sense a proton or a neutron is a fundamental particle; a particle without any substructure. They were called hadrons from the Greek word hadros meaning bulky. In late 1960's Gabriele Veneziano, an Italian scientist, to explain the zoo of particles proposed string theory as a model for hadrons. The model was abandoned when scientists discovered the quark structure of hadrons. In 1980's the model was revived as a quantum theory of gravity and much more; the theory of everything. Why is it called the theory of everything? All the activities in nature are governed by only four kinds of forces; (i) Gravitational force which make the Earth rotate about the Sun as well make an apple fall to the ground, (ii) Electromagnetic force which make an electron rotate about the nucleus, (iii) Strong nuclear force which keep the protons and neutrons bound within a nucleus, and (iv) Weak nuclear force which is responsible for the nuclear beta decay. Apart from quantizing gravity, string theory envisages unification of all the four forces. From ages, scientists are trying to unify the forces. In 1865, the first step was taken by James Clerk Maxwell when he unified electricity and magnetism. 1920 onwards, unification of gravity and electromagnetism (strong and weak forces were yet to be discovered) was the cherished dream of Einstein. He spent his later life in the vain attempt to unify electromagnetism with gravity. Next progress towards unification came in 1968, when three physicists, Sheldon Lee Glashow, Abdus Salam and Steven Weinberg were awarded Nobel Prize for their "Electroweak Theory" unifying electromagnetic force with the weak force. Later, an attempt was made to unify strong and electroweak interaction. The theory called Grand Unified theory, however, was not a success. It predicted decay of the proton, which was not observed experimentally. The string theory is called the theory of everything because it promises to fructify the long cherished dream of the scientists, to have one theory for all the fundamental forces of nature.

The basic idea of string theory is simple; fundamental or elementary particles are no longer point particles, rather they are string-like, have a dimension of length. Undoubtedly, the length scale is very small, of the order of Plank length ~ $10^{-33}$ cm. I will not go into the details, but in this theory, all the fundamental particles are nothing but different modes of vibration of the tiny strings. Trouble started when one applies Quantum mechanics and Relativity, the two pillars of the modern science.

(i) In our ordinary 4-dimensional world (with three spatial dimensions and one temporal dimension), string theory is not consistent with quantum mechanics and relativity. The theory is consistent with quantum mechanics and relativity either in 10 dimensions or in 26 dimensions.

(ii) The theory also requires a kind of symmetry called supersymmetry. Physicists endow each elementary particle with a characteristic quantity called spin. Spin can be either half-integer (1/2, 3/2...) or integral (0, 1, 2...) but not in between; say 1/3. A half-integer spin particle is called Fermion (after the Italian scientist Enrico Fermi) and an integral spin particle is called Bosons (after the Indian scientist Satyendra Nath Bose). In supersymmetry, every fundamental particle has a superpartner, i.e., for every fermion type of particle, there is a bosonic type of particle and the vice versa. There is an additional problem; supersymmetry can be incorporated in five different ways, giving rise to five types of string theory. Later, it was discovered that the five versions of the superstring theory are solutions of an 11-dimensional theory called M-theory. What "M" stands for is uncertain. M can stand for the membrane, because, in one way or other, theory contains surfaces or membranes.

Both the extra-dimensions and supersymmetry are not observed in nature. String theorists found a way out. Using a mathematical trick, extra-dimensions were compactified or curled up in a small circle to make them unobservable. From the stability criterion, the trick can be applied only on certain kind of space called Calabi-Yau space, a six-dimensional mathematical space named after Eugenio Calabi, an Italian-American mathematician and Shing–tung Yau, a Chinese-American mathematician. Unfortunately for the string theorists, Calabi-Yau space is not unique, there are hundred thousands of Calabi-Yau space, on each of which the extra-dimensions can be curled up, resulting into hundred thousands of string theories, each of which is different. The problem of supersymmetry was circumvented by postulating it to be symmetry of nature at very high energy, e.g. at the early Universe. The symmetry is broken at lower energy, as in our present Universe. String theorists are uncertain about the mass of the superpartners, but if the symmetry breaks at the energy scale of E, the mass of the super-particles are expected to be of the order of the symmetry breaking energy scale. Now, if supersymmetry is symmetry of nature, one age-old problem in physics known as the hierarchy problem is also solved. There are several ways to pose the hierarchy problem. One simple way to pose the problem is why the strongest and weakest forces of nature differ by a factor of $10^{38}$? Theoretically, supersymmetry can provide for a solution to the problem if the symmetry breaks at a scale of 1000 GeV. If supersymmetry breaks at the energy scale of 1000 GeV, the masses of the super particles are expected to be of the same order. They should be produced in the Large Hadron Collider experiments at CERN, where two protons can be collided at an enormous energy of 13000 GeV. However, till today, there is no evidence of super-particles in Large Hadron Collider experiments. While there are encouraging results, string theory is far from complete and till today, the theory has no definite prediction that can be tested. There are several criticisms against it. One of the severest critics of string theory was Richard Feynman. He thought that the theory is crazy and is in the wrong direction. When asked why he did not like the theory, he replied,

*"I don't like that they're are not calculating anything. I don't like that they don't check their ideas. I don't like that for anything that disagrees with an experiment, they cook up an explanation- a fix-up to say "Well, it still might be true." For example, the theory requires ten dimensions. Well, maybe*

*there's a way of wrapping up six of the dimensions. Yes, that's possible mathematically, but why not seven? When they write their equation, the equation should decide how many of these things wrapped up, not the desire to agree with experiment. In other words, there's no reason whatsoever in superstring theory that is isn't eight of the ten dimensions that get wrapped up and that the result is only two dimensions, which would be completely in disagreement with experience. So the fact that it might disagree with experience is very tenuous, it doesn't prove anything; it has to be excused most of the time. It doesn't look right."*

The questions raised by Feynman are yet to be answered.

**4. String theory under the lens of the demarcation criteria**

In the positivistic approach, to be scientific, statements or theory should be verifiable. Let us consider the string theory statement on extra-dimensions: our Universe has more than four dimensions. Is it verifiable? No. By choice, the theory has been made to be beyond the experimental verification. The extra-dimensions are too small to be revealed unless we can build an accelerator to accelerate particles to $1.22 \times 10^{19}$ GeV, inconceivable even in foreseeable future. Is string theory statement about supersymmetry being a symmetry of nature at high energy verifiable? In principle, yes. If supersymmetry is symmetry of nature at high energy and breaks at lower energy scale, then super particles will be produced in collisions of particles at energy equal to the symmetry breaking energy scale or more. From theoretical considerations, supersymmetry is expected to break around 1000 GeV. In the Large Hadron Collider, experiments have been performed by colliding particles at 13000 GeV, yet no evidence for the super particles could be obtained. The string theory statement about supersymmetry could not be verified. Also, since the theory has no definite prediction; we have no scope of verifying any prediction from string theory. Overall, string theory fails miserably the verifiability criterion of science.

Can the theory be falsified? As it cannot be verified, it cannot be falsified as well. Again, by choice, the theory is constructed such that it is beyond the falsification criterion. For example, can we falsify the basic premise of the string theory that the natural world has more than 4-dimensions? No, in foreseeable future we will not be able to build accelerators to probe the tiny extra-dimensions. Can we falsify string theory statement about supersymmetry being a symmetry of nature at high energy? To falsify the statement we have to prove that super particles do not exist. Since string theory does not have definite prediction about the mass of the super particles, the statement cannot be falsified. Practitioners of the theory can always argue that they too heavy to be produced in present day accelerators. Also, since the theory has no definite prediction; we have no scope of falsifying any prediction of the theory. In Popper's falsification criterion string theory is not science.

What about Kuhn's puzzle solving criterion? According to Kuhn, within an accepted paradigm of worldview, experiences and observations throw some anomaly, some puzzle, which normal science try to solve. Since the paradigms of extra-dimensions or supersymmetry are not accepted paradigms we cannot say that there is any puzzle in string theory. Is it possible that string theory is a revolutionary science? In a sense it is revolutionary science. It is trying to change not one but three existing paradigms of the accepted world view. Present paradigm is that a fundamental or elementary particle is a point particle. String theory wants to change that paradigm to the paradigm of elementary particles with the dimension of length. Present paradigm is that the world is 4-dimensional. String theory wants to change the paradigm to a paradigm of 10-dimensional World. In the present paradigm supersymmetry is not a symmetry of nature. String theory wants to change that paradigm also. However, according to Kuhn, revolutionary science should satisfy Popper's criterion of falsification and as discussed above, string theory cannot be falsified. Moreover, revolutionary science emerges when anomalies in normal science become progressively larger and larger such that they can no longer be accommodated within the existing paradigm. String theory, however, emerged from the elusive lure of the unification of fundamental forces. Indeed, without the lure of unification, it is doubtful whether the theory would have been as popular as it is.

Indeed, it is strange that the theory was not abandoned when consistency with relativity and quantum mechanics required extra-dimensions, a world of dimension 10 or 26. Instead of abandoning the theory, the practitioners took the route to detach the theory from physical reality by making the extra-dimensions small and unobservable. Continuation of the unphysical theory gave rise to a bizarre situation like multiverse; with hundred thousands of Universes, each with its own initial conditions and our universe being only one among the hundred thousands of Universes. The situation is more akin to fiction than science, because we will never have the opportunity to verify the concept, intrinsically, the concept is beyond verification.

David Hume [4] had a simple way to distinguish between science and non-science. In 'An Enquiry concerning Human Understanding' he wrote;

*"If we take in our hand any volume; of divinity or school metaphysics, for instance; let us ask, Does it contain any abstract reasoning concerning quantity or number? No. Does it contain any experimental reasoning*

*concerning matter of fact and existence? No. Commit it then to the flames: for it can contain nothing but sophistry and illusion."*

Statements or theory, unless reasons about number or quantity, unless are supported by experiments, are a mere sophistry and illusion. Undoubtedly, Hume would have committed to flame, any volume on string theory.

**5. Conclusion**

In conclusion, we have examined the demarcation problem and opined that for an activity to be deemed as science a set of criteria are needed to be satisfied. The set of criteria are;

(i) Positivistic verifications: theories or statements can be verified;
(ii) Karl Popper's falsification: theories or statements can be falsified;
(iii) Kuhn's puzzle-solving: within a given paradigm, existence of anomaly or puzzle at any time;

The set of criteria, when applied to string theory, led us to conclude that the theory cannot be classified as science. Statements of the theory cannot be verified nor falsified. Also, since the theory is trying to change existing paradigms, it does not have puzzles in the usual sense.